\documentstyle[pra,aps,twocolumn,epsf]{revtex}
\renewcommand{\vec}[1]{\mbox{\boldmath$#1$}}

\begin{document}
\draft
\preprint{cond-mat}
\twocolumn[\hsize\textwidth\columnwidth\hsize\csname @twocolumnfalse\endcsname

\title{
Crucial Effects of Intramolecular Charge Distribution on the Neutral-Ionic Transition of Tetrathiafulvalene - {\em p}-Chloranil
}

\author{
Tohru Kawamoto$^1$, Takako Iizuka-Sakano$^1$, Yukihiro Shimoi$^{1,2}$, and Shuji Abe$^1$
}
\address{
$^1$
Nanotechnology Research Institute, 
National Institute of Advanced Industrial Science and Technology, 
1-1-1 Umezono, Tsukuba 305-8568, Japan \\
$^2$
Institute for Molecular Science, Myodaiji, Okazaki 444-8585, Japan 
}
\date{\today}
\maketitle

%
%

\begin{abstract} 
We calculated the Madelung energies of both the ground state and excited states in tetrathiafulvalene - $p$-chloranil (TTF-CA) by taking into account intramolecular charge distribution. 
The distribution is found to be significant in the neutral-ionic (NI) transition. In the ionic phase, the Madelung energy depends more strongly on the intermolecular distance perpendicular to the $\pi$-stacking chains than on that along the chains. 
This indicates that simple single-chain models neglecting interchain electrostatic coupling are not adequate. 
The gain of the Madelung energy due to dimerization is concluded to be small compared with the other structural changes. 
We also calculated the formation energy of excited state domains, which appear in the primary process of the phase transition. 
A one-dimensional excited domain has the smallest energy among the possible domains with the same number of molecules when the domain is small, and the energy per molecule is considerably reduced in increasing the domain size. 
It is consistent with the experimental suggestion that a large number of excitations were generated by only one photon. 

\end{abstract}
\vskip2pc]

\narrowtext

\section{Introduction}
%
%

Controlling properties of materials by various external stimuli has been attracting attention from the viewpoint of fundamental researches and applications.
An organic compound tetrathiafulvalene - $p$-chloranil, TTF-CA, is one of the typical model materials\cite{Torrance81PRL1,Torrance81PRL2,Lemeecailleau97PRL,Koshihara90PRB,Koshihara99JPCB}, where a phase transition accompanied by the change of dielectric response\cite{Okamoto91PRB} is induced by applying pressure, or varying temperature, or irradiation of light. 
Torrance {\em et al.} first discovered the pressure-induced neutral-ionic (NI) phase transition of TTF-CA\cite{Torrance81PRL1}, followed by the demonstration of the temperature-induced one\cite{Torrance81PRL2}.
At high temperature (or at low pressure), both TTF and CA molecules are nearly neutral in their electronic charges (N-phase). 
On the other hand, at low temperature (or at high pressure), they are quasi-ionic  (I-phase) due to electron transfer from TTF to CA. 
The degrees of charge transfer $\rho$ were estimated to be about 0.3 in the N-phase and 0.7 in the I-phase\cite{Jacobsen83JCP}.
The photo-induced changes relevant to the NI transition were also reported\cite{Koshihara90PRB,Koshihara99JPCB,Suzuki99PRB}. 
The I-phase is partially converted to the N-phase by irradiation of laser pulse with photon energy 2.0-2.5 eV.
It was reported that about 300 pairs of TTF-CA molecules were converted by one photon\cite{Koshihara99JPCB}, suggesting that this phenomenon involves a cooperative effect with large nonlinearity. 

\begin{figure}
\begin{center}
  \leavevmode
  \epsfxsize=80mm
  \epsfbox{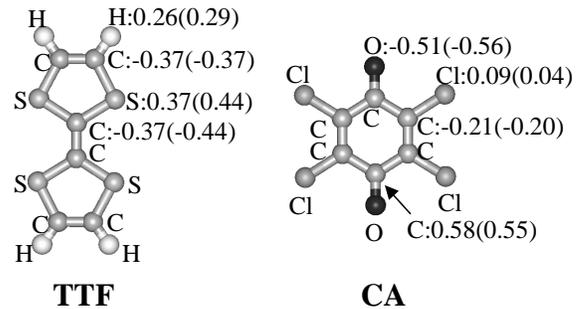}
\end{center}
\caption{
Structure of TTF and CA molecules. 
The calculated charge distributions on each atom are also shown for $\rho$=0.3, and for $\rho=0.7$ in parentheses.
}
\label{fig:intra_mol}
\end{figure}

%
%
The planer TTF and CA molecules shown in Fig.~\ref{fig:intra_mol} form mixed-stacking chains along the $a$-axis in the crystal\cite{Lecointe95PRB}. 
The temperature-induced phase transition as well as the pressure-induced one is understood in terms of changes in electrostatic Madelung energy by lattice expansion or contraction.
A dimerized structure was observed in the I-phase\cite{Lecointe95PRB}. 
The origin of the dimerization was also ascribed to the electrostatic interaction in previous theoretical works\cite{Avignon86PRB,Luty95APPA,IizukaSakano96JPSJ,Huai00JPSJ}. 

%
%
In most of the previous works, TTF-CA was treated as a quasi one-dimensional chain along the mixed-stacking  $a$-axis with weak or no interchain coupling with respect to the Madelung interaction as well as to the overlapping of the $\pi$-orbitals \cite{Koshihara99JPCB,Huai00JPSJ,Yartsev98OM}. 
That is, the electrostatic interaction between chains was assumed to be much smaller than that within the chain \cite{Nagaosa86SSC}. 
This assumption was justified in the approximation where each molecule was replaced by a point charge at its center of mass. 
Hereafter this is called the point molecule approximation. 
The nearest neighbor molecules of TTF or CA are the same along the $b$-axis and the different along the $\vec{b}+\vec{c}$ direction\cite{Hubbard81PRL}. 
As a result, a repulsive interaction along the $b$-axis and an attractive one along the $\vec{b}+\vec{c}$ direction are expected to cancel out each other\cite{Nagaosa86SSC}. 
However, this idea is not consistent with the change observed in the lattice constant\cite{Lecointe95PRB}. 
The lattice constant $b$ becomes small at the NI transition with lowering temperature.
This implies an attractive interaction along the $b$-axis. 
In contrast, a change in the lattice constant $a$ is very small, although a magnitude of the interaction changes at the transition. 

%
%
In the present paper, we will demonstrate the inadequacy of the point molecule approximation. 
In the calculation of Madelung energy, we take account of the charge distribution on atoms inside each molecule derived with an ab initio quantum chemical method\cite{Metzger85JACS}. 
We call this approach the point atom approximation. 
Another purpose of this paper is the elucidation of the primary process of the phase transition. 
We estimate the formation energy of an excited-state domain, {\em e.g.} a neutral domain in the ionic phase. 
This will provide information about the growing process of the excited domain and inter-domain interactions. 

This paper is organized as follows: 
The charge distribution of each molecule calculated with an ab initio technique is presented in Section II. 
In section III, the Madelung energy of the ionic phase is calculated with various modifications of the crystal structure to show the inadequacy of the point molecule approximation. 
Section IV deals with the formation energy of the excited-state domain. 
Finally, we give concluding remarks in Section V. 
\section{Charge distribution inside molecules}
%
%
%
We calculated the intramolecular charge distribution in each molecule using the ab initio quantum chemical method\cite{Schmidt93JCC}. 
The calculations were performed both for neutral and monovalent states. 
The molecular structures were extracted from the crystal structure of TTF-CA determined at 90 K, in which the molecules have the inversion symmetry\cite{Lecointe95PRB}.
The restricted and unrestricted Hartree Fock methods were respectively used for the neutral and monovalent states with the 6-31G$^*$ basis function set. 
The charge density on each atom was calculated with the Mulliken population technique. 
The obtained density is consistent with the previous studies\cite{Metzger85JACS,Katan97SSC}. 
The charge distribution at an observed fractional charge $\rho$ is evaluated by a linear interpolation between the complete neutral and monovalent states. 

The results are shown in Fig.~\ref{fig:intra_mol}. 
The large electronic polarization is found on both the molecules. 
For example, the carbon atom next to an oxygen atom in the CA molecule has a large positive charge even if the total charge of the molecule is negative. 
The derived intramolecular charge distribution is used for the point atom approximation in the following sections. 

\section{Structure dependence of Madelung energies}
%
%
%
In this section, we report the Madelung energy, $E_{\rm M}$, of the ionic phase of TTF-CA using the Ewald method. 
We examine how the electrostatic energy changes with the following deformations of the crystal structure: the variation of the lattice constants, the molecular rotation around its center of mass, and the dimerization along the $a$-axis. 
The intramolecular atomic coordinates and the charge distribution are fixed because intramolecular deformations are smaller than intermolecular ones. 
This means that only the spacing among molecules is varied. 

\begin{figure}
\begin{center}
  \leavevmode
  \epsfxsize=80mm
  \epsfbox{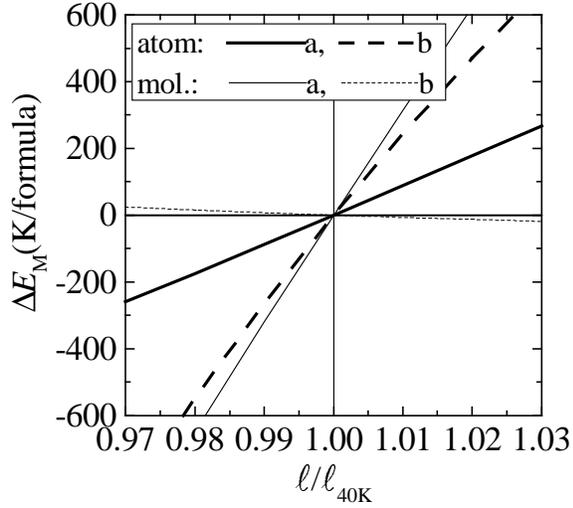}
\end{center}
\caption{
Shifts of $E_{\rm M}$ due to the lattice constant variation. 
The abscissa axis represents each lattice constant normalized by that
observed at 40 K, $\ell_{\rm 40K}$[9]. 
The results using the point atom approximation and with the point molecule approximation are represented by `atom' and `mol.', respectively (see text). 
}
\label{fig:madelung_vs_abc}
\end{figure}

%
%
Figure~\ref{fig:madelung_vs_abc} shows the variation of $E_{\rm M}$ against the uniaxial deformation along the $a$ and $b$ axes from the crystal structure observed at 40 K\cite{Lecointe95PRB}. 
The results are quite different between the two approximations. 
The lowering of $E_{\rm M}$ due to the contraction of the lattice constant $a$ is suppressed in the point atom approximation. 
As for the case of the $b$-axis, the slope of $E_{\rm M}$ against lattice constant is different in the two approximations. 
In the point molecule approximation, the slope is slightly negative, which agrees with the previous proposal of the repulsive electrostatic interaction along the $b$-axis\cite{Nagaosa86SSC,Hubbard81PRL}. 
In contrast, in the point atom approximation, $E_{\rm M}$ is decreases in contracting along the $b$-axis. 
In this approximation, the attractive interaction is stronger along the $b$-axis than along the $a$-axis, which is consistent with the observation that the lattice constant along the $b$-axis is much shortened at the phase transition temperature in going from the N-phase to the I-phase\cite{Lecointe95PRB}. 
With varying the lattice constant along the $c$-axis, we found a tendency similar to the case of the $b$-axis. 

\begin{figure}
\begin{center}
  \leavevmode
  \epsfxsize=80mm
  \epsfbox{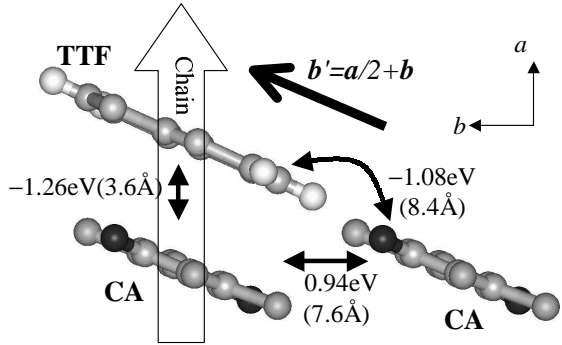}
\end{center}
\caption{
Molecular alignment within the $ab$-plane. 
The intermolecular electrostatic energy is shown for each pair. 
The distances between their centers of mass are also indicated along in parentheses. 
}
\label{fig:crystal_ab}
\end{figure}

%
%
Figure~\ref{fig:crystal_ab} shows the electrostatic energies between neighboring molecules in the $ab$-plane. 
The attractive energy between TTF and CA aligned along $\vec{a}/2+\vec{b}$, defined as $b^\prime$-axis, is found to overcome a repulsive one between CA's along the $b$-axis, while the intermolecular distance measured in terms of the center of mass of each molecule is larger in the former pair than in the latter. This indicates the inapplicability of the point molecule approximation. 
The discrepancy is mainly caused by the hydrogen bond along the $b^\prime$-axis. One of the hydrogen atoms of TTF is very close to one of the oxygen atoms of CA. This hydrogen bond makes a large contribution to the interchain attractive interaction. 

\begin{figure}
\begin{center}
  \leavevmode
  \epsfxsize=80mm
  \epsfbox{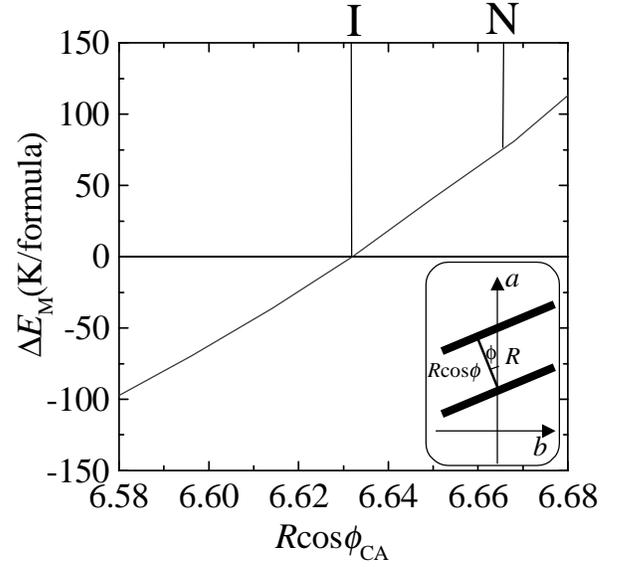}
\end{center}
\caption{
Shift of $E_{\rm M}$ due to the molecular rotation from the structure at the I-phase (40 K), where $R$ and $\phi_{\rm CA}$ represent the distance between the center of mass of neighboring CA molecules along the $a$-axis, and the molecular inclination angle for CA molecules defined in the inset, respectively. 
The angle 'I' and 'N' indicate the inclination angles $\phi_{\rm CA}$ observed at the I-phase (40 K) and at the N-phase (90 K)[9]. 
}
\label{fig:rotation}
\end{figure}

%
%
The inclination of the molecular planes against the stacking axis also has some influences on $E_{\rm M}$. 
At the NI transition, the molecules rotate mainly within the $ab$-plane. 
We define the inclination angle $\phi$ as shown in the inset of Fig.~\ref{fig:rotation}. 
The angle $\phi$ of the CA molecule ($\phi_{\rm CA}$) in the I-phase (40 K) is found to be larger by 1.4 degree than that in the N-phase (90 K), and found that $\phi$ of the TTF molecule ($\phi_{\rm TTF}$) in the I-phase is also larger by 0.7 degree.
These rotations alter the effective intermolecular distance  $R\cos\phi$ in the inset of Fig.~\ref{fig:rotation}. 
The dependence of $E_{\rm M}$ on $R\cos\phi_{\rm CA}$ is shown in Fig.~\ref{fig:rotation}. 
The change of the angle of TTF is set to be half that of CA to match the observed changes of $\phi_{\rm CA}$ and $\phi_{\rm TTF}$. 
The other geometrical parameters are fixed to the crystal structure determined at 40 K. 
$E_{\rm M}$ is an increasing function of $R\cos\phi_{\rm CA}$. 
This means that the Madelung energy is stabilized by the decrease of the effective intermolecular distance. 
The energy difference due to the molecular rotation at the NI transition is 80 K/formula, which is nearly the same magnitude as the effect of the contraction of the lattice constant $a$ by 1\% as shown in Fig.~\ref{fig:madelung_vs_abc}. 
This implies that the enlargement of an attractive interaction along the $a$-axis in the I-phase makes the effective distance between molecules to be short instead of the contraction of the lattice constant. 

%
%
The dimerization observed in the ionic phase has often been considered as an essential feature of the NI transition\cite{Luty95APPA,Huai00JPSJ}. 
We also calculated the stabilization energy due to the dimerization. 
The gain of the Madelung energy is 2.3 K/formula for a displacement of CA by 1 \% along the $a$-axis, which is comparable to the observed displacement in the ionic phase. 
The stabilization energy is much smaller than the energy changes due to the other deformations studied in this paper, as well as than the observed phase transition temperature, 81 K.  

\section{Formation energy of the excited domain}
\begin{figure}
\begin{center}
  \leavevmode
  \epsfxsize=80mm
  \epsfbox{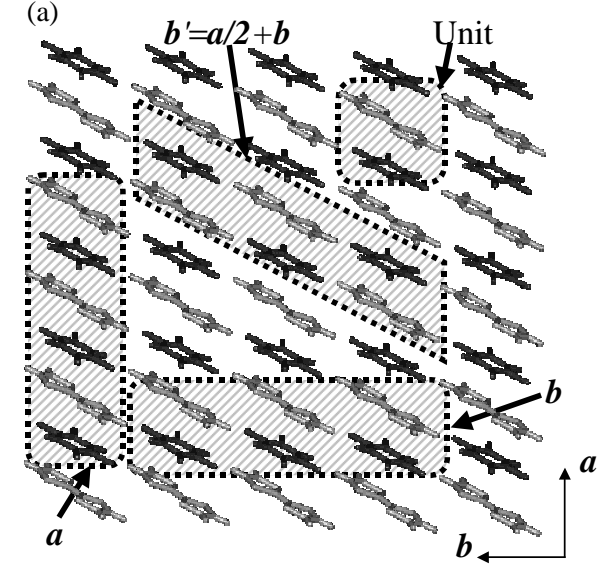} 

  \leavevmode
  \epsfxsize=80mm
  \epsfbox{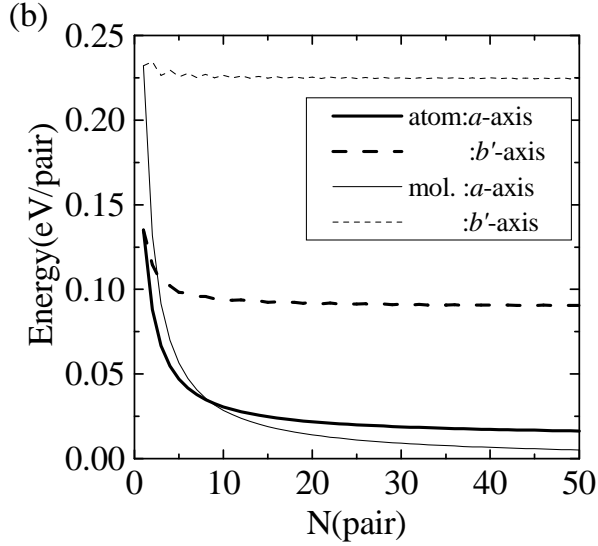}
\end{center}
\caption{
(a) Examples of the structure of 1D domains.
A constituent unit of the excited domain is represented by 'Unit'. 
(b) The formation energy of the 1D domains along each axis. 
The results using the point atom approximation and the point molecule approximation are represented by `atom' and `mol.', respectively (see text). 
}
\label{fig:ex_eng_string}
\end{figure}

%
%
In this section, we consider an excited-state domain in the ground state, {\em e.g.} a neutral domain in the ionic phase, in connection with the photoinduced NI transition. 
We assume that a pair of excited TTF-CA molecules along the $a$-axis is a constituent unit of the domain, because the transfer of a $\pi$ electron is likely to occur between TTF and CA along the stacking axis. 
The constituent unit and some examples of the domain are given in Fig.~\ref{fig:ex_eng_string}(a). 
We calculated the formation energy $E^{\rm EX}$ of the domain in an extended scheme of the point atom approximation. 
Here we show the result with the crystal structure at 90 K for simplicity and for avoiding the influence of the dimerization. 
$E^{\rm EX}$ is divided into two parts, 
\begin{eqnarray}
E^{\rm EX} = E_{\rm M}^{\rm EX} 
	+ N_{\rm ex}\Delta E_{\rm u} ,  \label{eq:esp}
\end{eqnarray}
where $N_{\rm ex}$ represents the number of the constituent units in the domain. $E_{\rm M}^{\rm EX}$ originates from the Madelung energy described as follows:
\begin{eqnarray}
E^{\rm EX}_{\rm M} &=& ( E_{\rm in}^{\rm EX} + E_{\rm env}^{\rm EX} )
                     - ( E_{\rm in}^{\rm GS} + E_{\rm env}^{\rm GS} ), \\
E_{\rm in}^{\alpha} &=& \frac{1}{2} \sum_{i \neq j \in D} 
				\frac{q_i^{\alpha} q_j^{\alpha}} {R_{ij}}, \\
E_{\rm env}^{\alpha} &=& \sum_{i\in D,j\notin D} 
            \frac{q_i^{\alpha} q_j^{\rm GS}} {R_{ij}},\alpha={\rm GS, 
or\hspace{2mm} EX},
\end{eqnarray}
where GS and EX indicate the ground state and the excited state of the constituent unit, respectively. 
$E_{\rm in}^{\alpha}$  denotes the Madelung energy in the excited domain, and $E_{\rm env}^{\alpha}$ is that between the domain and the surrounding ground state.
Group $D$ represents the set of atoms in the excited domain, and $i$ and $j$ specify atoms. 
$q_i^\alpha$ represents the charge of the $i$-th atom in the $\alpha$-state, and $R_{ij}$ means the distance between the $i$-th and $j$-th atoms. 
%
%
The second term of $E^{\rm EX}$ in Eq.\ref{eq:esp} represents all the contributions other than the Madelung energy, for example, electron affinity, ionization potential, and electron transfer energies. 
$\Delta E_{\rm u}$ is defined as $E_{\rm u}^{\rm EX} - E_{\rm u}^{\rm GS}$, where $E_{\rm u}^\alpha$ represents those contributions in the $\alpha$ state per constituent unit. 
Instead of calculating $E_{\rm u}^\alpha$ directly, we assumed that the N- and I-phases are degenerate in energy.  
In this case, $\Delta E_{\rm u}$ is equal to the difference of the Madelung energies per unit between the two phases. 
In the expression of $E_{\rm M}^{\rm EX}$, we find with a straightforward calculation that the formation energy does not change by interchanging {$q_i^{\rm GS}$} and {$q_i^{\rm EX}$}. 
This means that the N-domain in the I-phase has the same energy as the I-domain of the same shape in the N-phase. 

%
%
We investigated $E^{\rm EX}$ of the one-dimensional (1D) domains with various shapes, specifically along the $a$, $b$, $b^\prime$, and $c$-axes with the number of pairs ranging from 1 to 50.  
Figure~\ref{fig:ex_eng_string}(b) shows $E^{\rm EX}$ of 1D domains along the $a$ and $b^\prime$-axes, as a function of the domain size. 
For comparison, the results using the point molecule approximation are also shown. 
The formation energy is found to be sensitive to the direction of the domain growth. 
It is the lowest in the direction along the $a$-axis among all the types; that for the 50 excited pairs along the $a$-axis is at 0.016 eV per pair, which is only 1/8 as large as that of a single excited pair. 
Such a drastic reduction of the formation energy accords with the experimental suggestion that a large number of excited pairs are created by a single photon in the photoinduced change\cite{Koshihara99JPCB}. 
The formation energy for the 1D domain along the $b^\prime$-axis is much reduced by taking account of the molecular structure as for the comparison between the two approximations. 
This indicates, again, that the interchain coupling is much stronger in the real situation than in the point molecule approximation. 

%
%

\begin{figure}
\begin{center}
  \leavevmode
  \epsfxsize=80mm
  \epsfbox{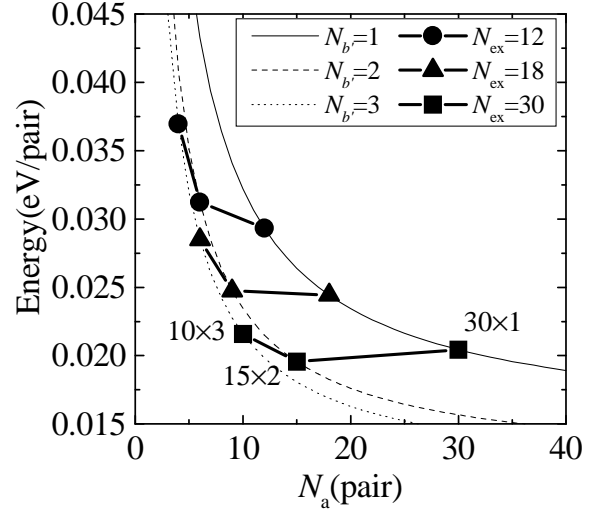}
\end{center}
\caption{
Formation energy of two-dimensional excited domain in the $(a,b^\prime)$ plane, where $N_a$, $N_{b^\prime}$, and $N_{\rm ex}(\equiv N_a \times N_{b^\prime}$) represents the number of TTF-CA pairs along the $a$-axis in the domain, that along $b^\prime = a/2+b$, and total number of pairs in the domain. 
Domain sizes $N_a \times N_{b^\prime}$ explicitly indicate for $N_{\rm ex}$=30.
}
\label{fig:ex_eng_2d}
\end{figure}

$E^{\rm EX}$ of the two- or three-dimensional excited domains were also calculated. 
In Fig.~\ref{fig:ex_eng_2d}, we show the results for two-dimensional domains within the $(a,b^\prime)$ plane. 
The excited domain has a shape of parallelogram consisting of $N_{\rm ex}$ excited pairs spanned by $N_a$ and $N_{b^\prime}$ units along the $a$ and $b^\prime$-axes.
It is found that the excited domain with $N_{b^\prime}$ = 1, {\em i.e.} the 1D domain along the $a$-axis, is the most stable among the domains of the same size in the case of $N_{\rm ex}$=12 or 18. 
This remains true as long as the number of excited pairs is less than twenty. 
This suggests that excited domains easily grow along the $a$-axis in the primary stage of the phase transition. 

\section{Concluding Remarks}
%
%
Here we comment on the electron transfer interaction that has been neglected in our calculations. 
Since the transfer integral of $\pi$ electrons between chains is considered quite small, the essential role of the attractive electrostatic interaction between chains would be unchanged even if we take account of the itinerancy of electrons. On the other hand, the electron transfer is considered to have some influence on the electronic structure within chains.
The fractional degree of the charge transfer that we assumed in this paper is ascribed to this interaction. 

The growth process of the excited domains in chains will also be affected. 
As well as the electrostatic interaction, the electron transfer is expected to accelerate the domain growth along the mixed-stacking chain. 
With the intermolecular electron transfer energy, the dynamics of the photoexcited state has been investigated by means of a one-dimensional modified Hubbard model\cite{Koshino00PRB}, in which the electrostatic interaction between molecules was neglected. 
For more quantitative description of the dynamics, it is necessary to develop a theory that includes both the electrostatic interaction with the intramolecular charge distribution and the electron transfer interaction. 

%
%
Next, we discuss the driving force of the photoinduced phase transition. 
We have calculated the effect of the lattice constant expansion or contraction on the Madelung energies. 
The 0.7 \% contraction of the lattice constant $b$, which was the observed difference between 40 K and 90 K, lowers $E_{\rm M}$ about 20 meV per TTF-CA pair. 
The energy gains due to the molecular rotation within the $ab$-plane and the dimerization along the $a$-axis is about 7 meV and 0.2 meV, respectively. 
On the other hand, the formation energies $E^{\rm EX}$, derived without any lattice relaxations, of an excited TTF-CA pair and the excited domain of 50 pairs are respectively 135 meV and 16 meV per pair. 
Such a drastic reduction of the formation energy is caused by the electrostatic interaction.  
The change in the Madelung energies is likely to be essential in the primary process after the photoexcitation. 
The change of the lattice constant $b$ also becomes important in increasing the excited domain size, where $E^{\rm EX}$ becomes comparable to the lattice relaxation energy. 
The role of the lattice constant variation was discussed for the cooperative photoinduced phase transition of spin-crossover complexes\cite{Hauser92CPL,Koshino99JPSJ,Ogawa00PRL}. 
Similar discussion may also be required in TTF-CA. 

In summary, we calculated the electrostatic energies in both the ground and excited states of TTF-CA. 
It has been concluded that the point molecule approximation, where each molecule is replaced by a point charge at its center of mass, is inadequate for this system. 
For the ground state, we have demonstrated the crucial effect of interchain interactions, especially between TTF and CA molecules aligned in the \vec{a}/2 + \vec{b} direction. 
For the excited states, we found large nonlinearity in the size dependence of the formation energy of the excited-state domain. 
The growth of the domain will proceed along the $a$-axis in the primary process after the photoexcitation.  


\clearpage

\end{document}